\documentclass[prl,twocolumn, notitlepage,superscriptaddress, nofootinbib]{revtex4-1}

\usepackage{fancyhdr}
\fancyheadoffset[L,R]{\headrulewidth}
\renewcommand{\headrulewidth}{0.6pt}

\usepackage{cancel}
\usepackage[normalem]{ulem}

\usepackage{cancel}

\usepackage{graphicx}
\usepackage{tikz,pgf}
\usepackage{color}
\usepackage{array}
\usepackage{xcolor}
\usepackage[dvips]{epsfig}
\usepackage{epsfig}
\usepackage{enumerate}
\usepackage[latin1]{inputenc}
\usepackage[T1]{fontenc}
\usepackage[hang]{subfigure}
\usepackage{bbm, dsfont}
\usepackage{amsfonts,amsmath,amssymb,stmaryrd}
\usepackage{ulem}

\usepackage{bbold}

\usepackage{simplewick}

\newcommand{\bra}[1]{\langle #1 | \,}
\newcommand{\ket}[1]{\, | #1 \rangle}

\newcommand{\ga}{\ga}

\newcommand{\bl}{\begin{linenomath*}}
\newcommand{\el}{\end{linenomath*}}

\newcommand{\bea}{\begin{eqnarray}}
\newcommand{\eea}{\end{eqnarray}}
\renewcommand{\vec}[1]{\mathbf{#1}}

\newcommand{\be}{\hat b}
\newcommand{\bed}{\hat b^\dagger}
\renewcommand{\ga}{\hat\gamma}

\newcommand{\veck}{\mathbf k}

\definecolor{dgreen}{rgb}{0.0, 0.5, 0.0}

\usepackage{hyperref}

\begin{document}

\title{Quasiparticle approach to molecules interacting with quantum solvents}

\author{Mikhail Lemeshko}
\email{mikhail.lemeshko@ist.ac.at}
\affiliation{IST Austria (Institute of Science and Technology Austria), Am Campus 1, 3400 Klosterneuburg, Austria}
\affiliation{Kavli Institute for Theoretical Physics, University of California, Santa Barbara, CA 93106, USA}

\begin{abstract}

Understanding the behavior of molecules interacting with superfluid helium represents a formidable challenge and, in general, requires approaches relying on large-scale numerical simulations. Here we demonstrate that experimental data collected over the last 20 years provide evidence that molecules immersed in superfluid helium form recently-predicted angulon quasiparticles [Phys. Rev. Lett. 114, 203001 (2015)]. Most important, casting the many-body problem in terms of angulons amounts to a drastic simplification and yields effective molecular moments of inertia as straightforward analytic solutions of a simple microscopic Hamiltonian. The  outcome of the angulon theory is in good agreement with experiment for a broad range of molecular impurities, from heavy to medium-mass to light species. These results pave the way to understanding molecular rotation in liquid and crystalline phases in terms of the angulon quasiparticle.

\end{abstract}

\maketitle

Among its many peculiar properties, superfluid  $^4$He is quite averse to mixing with impurities which could serve as a microscopic probe of the superfluid phase. As a result, for several decades after the discovery of superfluidity by Allen, Misener, and Kapitza~\cite{AllenNat38, KapitzaNat38}, only macroscopic -- hydrodynamic -- properties of superfluid helium have been studied in the laboratory. In the 1990's, however, it was demonstrated that atoms and molecules  can be trapped in superfluid helium if the latter forms little droplets containing on the order of a thousand helium atoms~\cite{GoyalPRL92, HartmannPRL95, GrebenevScience98, ToenniesAngChem04, SzalewiczIRPC08}. Over the following years, trapping atoms, molecules, and ions inside superfluid helium nanodroplets -- sometimes called `nanocryostats' -- emerged as an important tool of molecular spectroscopy~\cite{ToenniesARPC98, CallegariJCP01, ToenniesAngChem04, ChoiIRPC06, SzalewiczIRPC08, StienkemeierJPB06,  MudrichIRPC14}. Such nanodroplets allow to trap single molecules in a cold environment ($\sim 0.4$ Kelvin), thereby isolating them from external perturbations. This allows to record spectra free of collisional and Doppler broadening, as well as to study species that are unstable in the gas phase, such as free radicals.

While superfluid helium  does not cause a substantial broadening of molecular spectral lines, it affects molecular rotation. In particular, molecules in superfluid helium nanodroplets acquire an effective moment of inertia, that is larger compared to its gas-phase value~\cite{CallegariJCP01, ToenniesAngChem04}. The relative magnitude of the effect increases from lighter to heavier species and is somewhat similar to renormalization of the effective mass for electrons interacting with a crystalline lattice~\cite{AppelPolarons, EminPolarons, PolaronsExcitons, Devreese15}.

Semiclassically, molecular rotation in helium can be rationalized within the `adiabatic following' model~\cite{GrebenevOCS, CallegariPRL99, CallegariPRL99err, LehmannJCP01, LehmannJCP02, PatelJCP03, ToenniesAngChem04, SzalewiczIRPC08}. There, it is assumed that the molecule induces a local density deformation (a `non superfluid shell') of helium which co-rotates along with the molecule, thereby increasing its moment of inertia. However, such a classical approach does not allow to get insight into the intriguing aspects of the problem arising from quantum many-body physics. Helium, on the other hand, represents a dense, strongly-interacting superfluid, where only a tiny fraction of $6-8\%$ forms a Bose-Einstein Condensate (BEC), even at zero temperature~\cite{LeggettQuantLiquids}. As a result, a detailed quantum mechanical understanding of molecular impurities in helium requires first-principles approaches based on extensive numerical simulations~\cite{SzalewiczIRPC08}. During last years, several  numerical studies, based mainly on path-integral, variational, and  diffusion quantum Monte-Carlo (MC), have been performed for molecules embedded in finite-size He$_n$ clusters with $n\lesssim 100$~\cite{BarnettHeSF6, BlumeJCP96, LeeHeSF6, KwonJCP96, KwonJCP00, PaesaniJCP01, MoroniPRL03, PatelJCP03, TangPRL04, PaesaniJCP04, MoroniJCP04,  ZillichHeC2H2, ZillichHeHCN,  PaesaniPRL05, ZillichJCP05, PaoliniJCP05, TopicJCP06,   VielJCP07,SkrbicJPCA07, MiuraJCP07, vonHaeftenHeCO, ZillichHeLiH, MarkovskiyJPCA09, RodriguesIRPC16}.

In this Letter we show that such an involved many-particle problem simplifies tremendously, if one assumes that molecules in helium droplets form angulons -- recently introduced quasiparticles consisting of a quantum rotor dressed by a field of many-body excitations~\cite{SchmidtLem15, SchmidtLem16, LemSchmidtChapter, Bikash16, Redchenko16, Li16, Yakaboylu16, Shepperson16}. The angulon theory is inherently many-body and describes interactions between a molecule and an \textit{infinite} number of helium atoms. Nevertheless, it still allows to derive the effective molecular rotational constants as simple analytic solutions of a microscopic Hamiltonian, and assign them a transparent physical interpretation. Moreover, the resulting agreement of the angulon theory with experiment provides a strong evidence for the angulon formation inside helium droplets.

We start from introducing the angulon Hamiltonian, which describes interactions of a rotating molecule with a bosonic bath~\cite{SchmidtLem15}:
\begin{equation}
\label{hamiltonian}
\begin{split}
 \hat{H} &= B \hat{\vec{J}}^2 + \sum\limits_{k\lambda\mu} \omega_k \hat{b}^\dag_{k\lambda\mu} \hat{b}_{k\lambda\mu} \\ 
 &+ \sum\limits_{k\lambda\mu} \sqrt{\frac{4 \pi}{2\lambda+1}} U_\lambda(k) ~ [Y^*_{\lambda\mu}(\hat{\theta},\hat{\phi})\hat{b}^\dag_{k\lambda\mu}+Y_{\lambda\mu}(\hat{\theta},\hat{\phi}) \hat{b}_{k\lambda\mu}],
 \end{split}
\end{equation}
where $Y_{\lambda\mu}(\hat{\theta},\hat{\phi})$ are spherical harmonics~\cite{Varshalovich},  $\sum_k \equiv \int dk$, and  $\hbar \equiv 1$. The first term of Eq.~\eqref{hamiltonian} corresponds to the rotational kinetic energy of the molecule, with $\hat{\vec{J}}$ the angular momentum operator. $B = 1/(2I)$ is the molecular rotational constant, where $I$ is the molecular moment of inertia. While the first term of Eq.~\eqref{hamiltonian} describes rotations of a linear rigid rotor, one can use it to describe an average kinetic energy of more complex molecules, such as symmetric and asymmetric tops~\cite{TownesSchawlow, BernathBook}, aiming to obtain an average renormalization of their rotational constants. Thus, the bare eigenstates of the impurity are given by the $(2L+1)$-fold degenerate levels $ | L,M\rangle$ with energies $E_L = B L(L + 1)$, where $L$ is the angular momentum quantum number, and $M$ its projection on the laboratory-frame $z$-axis. 
The second term of the Hamiltonian~\eqref{hamiltonian} represents the kinetic energy of the superfluid excitations (such as phonons and rotons), as given by the dispersion relation $\omega_k$. Here, the operators $\hat{b}^\dag_{k\lambda\mu}$ ($\hat{b}_{k\lambda\mu}$) are  creating (annihilating) a bath excitation with linear momentum $k=|\vec{k}|$, the angular momentum $\lambda$, and its projection $\mu$, onto the $z$-axis. These operators can be obtained from the spherical-harmonic expansion of the usual creation and annihilation operators,  $\bed_\veck$ and $\be_\veck$, defined in Cartesian space, see Refs.~\cite{SchmidtLem15, SchmidtLem16, LemSchmidtChapter} for details.

The last term of the angulon Hamiltonian~\eqref{hamiltonian} describes the interaction between the molecular impurity and the superfluid, where the coupling constants $U_\lambda(k)$ are proportional to the Legendre moments of the molecule-helium potential energy surface (PES) in  Fourier space. Note that the impurity-bath coupling explicitly depends on molecular angle operators, $(\hat \theta, \hat \phi)$, which makes Eq.~\eqref{hamiltonian} substantially different from, e.g., the Bose-polaron~\cite{Devreese15} or the spin-boson~\cite{LeggettRMP87} models. The Hamiltonian~\eqref{hamiltonian} was originally derived to describe an ultracold molecule interacting with a dilute BEC, where the coupling constants $U_\lambda(k)$ assume a simple analytic form~\cite{SchmidtLem15,LemSchmidtChapter}. In order to reproduce experimental data for a dense superfluid of $^4$He, however, we will approach Eq.~\eqref{hamiltonian} from a phenomenological perspective, by analogy with effective field theories of nuclear~\cite{MachleidtPR11} and condensed matter~\cite{FradkinFTCM} physics. 

Namely, we use a simple, one-parameter model to extract $U_\lambda (k)$ from the \textit{ab initio} PES calculations available in  the literature. First, we note that broadening of the spectral lines in superfluid helium~\cite{LehnigFD09} and solid \textit{para}-H$_2$~\cite{KatsukiPRL00} is dominated by rotational dephasing as opposed to decay. Therefore, we can assume that $U_\lambda$'s with even $\lambda$ play the main role, since they can lead to boson scattering which preserves molecular angular momentum. Furthermore, since for most molecules, the $\lambda=2$ channel is dominant~\cite{StoneBook13}, for the sake of simplicity we neglect the rest of the $U_\lambda$ terms.

Second, we assume that the coupling constant in the $\lambda=2$ channel can be approximated as:
\begin{equation}
\label{U2}
U_2 (k) = \Delta f(k)
\end{equation}
Here, the form-factor $f(k)$ is considered to be the same for all the molecular species, while the anisotropy parameter $\Delta$ depends on a particular molecule. Thus, the strength of the molecule-superfluid interactions can be quantified by the dimensionless parameter,
\begin{equation}
\label{gamma}
\gamma = B/\Delta,
\end{equation}
and the species with $\gamma < 1$ and $\gamma > 1$ belong to the strong-coupling and weak-coupling regimes, respectively.

We evaluate the anisotropy parameter $\Delta$ as:
\begin{equation}
\label{DeltaPar}
	\Delta  = \frac{\vert V_\text{eff}^\parallel - V_\text{eff}^\perp\vert}{2} \sqrt{\frac{5}{4\pi}},
\end{equation}
where $V_\text{eff}^\parallel$ and $V_\text{eff}^\perp$ are the effective molecule-helium interactions, derived from the  \textit{ab initio} PES calculations~\cite{ZangJCP14, KalemosJCP12, GutierrezHeI2, PetersonJCP05, HowsonHeOCS, ChangN2O, RanHeCO2, CallegariHeHCCCN, PatersonHeOH, HeijmenHeCO, KlosHeNO, TaylorHeLiH, MoszynskiHeHF, MurdachaewHeHCL, FernandezHeC2H2, HaradaHeHCN, DagdigianHeCH3, HodgesHeNH3, PatkowskiHeH2O, PackHeSF6, BarnettHeSF6, CalderoniHeCH4} as follows~\cite{sup}. For linear molecules, $V_\text{eff}^\parallel$ and $V_\text{eff}^\perp$ correspond to the effective molecule-helium interactions in the linear and T-shaped geometries, respectively. In most cases, the values of $V_\text{eff}$ were set to the average depths of the minima/saddle points in the corresponding configurations. If, for one of the configurations the PES  was purely repulsive, the corresponding $V_\text{eff}$ was set to zero, to reflect the fact that the helium density vanishes in this region. For symmetric and asymmetric tops (CH$_3$, NH$_3$, H$_2$O), $V_\text{eff}^\parallel$ was evaluated along the main molecular symmetry axis, while $V_\text{eff}^\perp$ along the direction perpendicular to it, laying within the mirror symmetry plane of the system. For the spherical-top molecules (SF$_6$, CH$_4$),   $\Delta$ was evaluated as an average anisotropy of PES minima which are not symmetry equivalent. In this case, $\vert V_\text{eff}^\parallel - V_\text{eff}^\perp\vert$ in Eq.~\eqref{DeltaPar} was replaced by $\sum_{i\neq j} \vert V_\text{eff}^{(i)} - V_\text{eff}^{(j)}\vert$, where $i,j$ label all non-equivalent minima of the PES~\cite{sup}.

 We would like to emphasise that we are quite aware of the fact that such  a one-parameter model provides a very rough approximation to the two-body interaction potential. However, as we can see below, it suffices to obtain a good agreement with experiment.

For molecules in helium droplets, the low-energy rotational spectrum is usually approximated as $E_{L} \approx B^\ast L(L+1)$, where $B^\ast$ is the effective rotational constant~\cite{ToenniesAngChem04}.
Let us first derive  $B^\ast$ from Eq.~\eqref{hamiltonian} in the strongly-interacting regime, $\gamma \ll 1$. Getting insight into this regime is inherently challenging, since it involves coupling molecular rotational angular momentum to angular momenta of, in principle, an infinite number of superfluid excitations. However, the solution can be drastically simplified by using a  canonical transformation recently introduced by Schmidt and Lemeshko~\cite{SchmidtLem16}:
\begin{equation}
\label{Transformation}
	 \hat{S} = e^{- i \hat\phi \otimes \hat \Lambda_z} e^{- i \hat\theta  \otimes \hat\Lambda_y} e^{- i \hat\gamma  \otimes\hat \Lambda_z} \\
\end{equation}
Here $(\hat\phi, \hat\theta, \hat\gamma)$ are the angle operators which act  in the  Hilbert space of the molecular rotor, and
\begin{equation}
\label{Lambda}
	 \hat {\vec\Lambda}=\sum_{k\lambda\mu\nu}\bed_{k\lambda\mu}\boldsymbol\sigma^{\lambda}_{\mu\nu}\be_{k\lambda \nu}
\end{equation}
is the total angular momentum operator of the superfluid excitations, acting in their corresponding Hilbert space. The matrices $\boldsymbol\sigma^{\lambda} \equiv \{\sigma^{\lambda}_{- 1}, \sigma^{\lambda}_{0}, \sigma^{\lambda}_{+1} \}$  fulfill  the $SO(3)$ algebra in the representation of angular momentum $\lambda$.
Thus, the  transformation operator of Eq.~\eqref{Transformation}  transfers the superfluid degrees of freedom into the frame co-rotating along with the molecule.

The transformation~\eqref{Transformation} brings the Hamiltonian~(\ref{hamiltonian}) to the following form~\cite{SchmidtLem16}:
\begin{multline}
\label{transH}
\hat{\mathcal{H}} \equiv \hat S^{-1} \hat H \hat S= B (\hat{\mathbf{L}} - \hat{\mathbf{\Lambda}})^2   + \sum_{k\lambda\mu}\omega_k \bed_{k\lambda\mu}\be_{k\lambda\mu} \\ + \sum_{k\lambda}  U_\lambda(k) \left[\bed_{k\lambda0}+\be_{k\lambda0}\right]
\end{multline}
 where $\hat{\mathbf{L}} = \hat{\mathbf{J}} + \hat{\mathbf{\Lambda}}$ is the \textit{total} angular momentum of the system, which acts in the rotating frame of the impurity and therefore obeys anomalous commutation relations~\cite{SchmidtLem16, LevebvreBrionField2, BiedenharnAngMom}.

In the limit of $\gamma \to 0$, the transformed Hamiltonian~\eqref{transH} can be diagonalized exactly, with the ground state for each $\ket{L M}$ given by:
\begin{equation}
\label{transHground}
	\ket{\psi_{LM}} =  e^{  \sum_{k \lambda}  \frac{ U_\lambda (k)}{\omega_{k}} \left( \be_{k \lambda 0} - \bed_{k \lambda 0}  \right)} \ket{0} \ket{LM}
\end{equation}
Note that since the boson coherent state of Eq.~\eqref{transHground} involves an infinite number of superfluid excitations, it would be quite challenging, due to the angular momentum algebra involved, to obtain this result starting from the non-transformed Hamiltonian, Eq.~\eqref{hamiltonian}.

Eqs.~\eqref{transH} and~\eqref{transHground} provide a transparent physical interpretation of molecular interactions with a  superfluid. For a slowly-rotating molecule, the superfluid coherent state of Eq.~\eqref{transHground} does not change upon molecular rotation. In a way, it can be thought of as a quantum formulation of the `nonsuperfluid helium shell' which rotates along with the molecule~\cite{GrebenevOCS, ToenniesAngChem04, SzalewiczIRPC08}. On the other hand, the effective molecular angular momentum, cf.\ the first term of Eq.~\eqref{transH}, is given by the difference between the total angular momentum of the system,  $ \hat{\mathbf{L}}$, and the superfluid angular momentum, $\hat {\vec\Lambda}$. Thus, the energy of a state with a given \textit{total} angular momentum $L$ is lower in the presence of a superfluid ($\hat {\vec\Lambda} \neq 0$) compared to a free molecule ($\hat {\vec\Lambda} = 0$), which leads to an effective renormalization of the rotational constant.

In the strong-coupling limit, the angular momentum of the superfluid is given by:
\begin{equation}
\label{ExpLambda}
	  \langle  \hat{\mathbf{\Lambda}}^2 \rangle \equiv \bra{\psi_{LM}} \hat{\mathbf{\Lambda}}^2 \ket{\psi_{LM}} = \sum_{k \lambda} \lambda(\lambda+1) \frac{ U^2_\lambda (k)}{\omega^2_{k}} 
\end{equation}

\begin{figure}[t]
\centering
\includegraphics[width=0.5\textwidth]{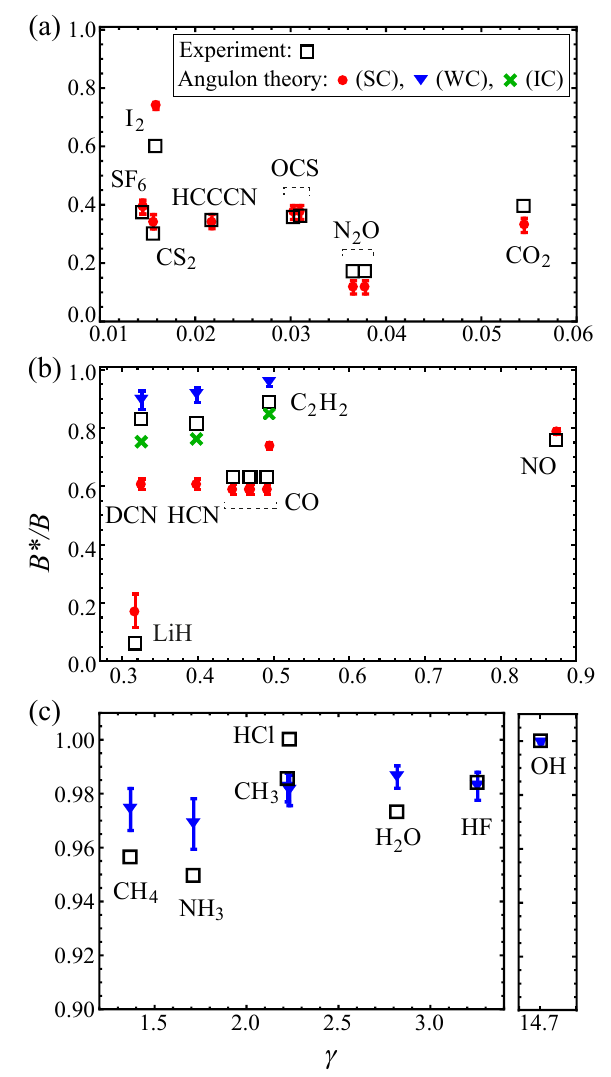}
\caption{Renormalization of the molecular rotational constant, $B^\ast/B$, as a function of the coupling parameter  $\gamma$. The panels correspond to (a) heavy molecules, (b) medium-mass molecules, and (c) light molecules. Experimental data from Refs.~\cite{ZillichI2, Shepperson16, AkinOjoJCP03, GrebenevOCS, NakautaCO2N2O,  ConjusteauHeHCN, ZillichHeHCN,  ZillichHeLiH, vonHaeftenHeCO, NautaHeC2H2, ZillichHeC2H2, vonHaeftenHeNO, SkvortsovHeHCl, NautaHeHF, RastonHeOH,  SlipchenkoHeNH3, MorrisonJPCA13, KuyanovHeH2O, HartmannHeSF6, LeeHeSF6, NautaHeCH4} (empty squares) are compared with the angulon theory in the strong-coupling regime, Eq.~\eqref{BstSC} with $\eta=[0.060 \pm 0.003 ]/$cm$^{-1}$ (red circles), and the weak-coupling regime,  Eq.~\eqref{BstPert} with $\xi=[0.0092 \pm 0.0028]/$cm$^{-1}$ (blue triangles). Green crosses show the intermediate-coupling interpolation between the strong- and weak-coupling theories.}
\label{fig:BBstar}
\end{figure}

In the first order, we can assume that $\hat{\mathbf{\Lambda}} \uparrow  \uparrow \hat{\mathbf{L}}$, i.e. $\hat{\mathbf{\Lambda}} = \alpha(L) \hat{\mathbf{L}}$, where the proportionality constant can be calculated as $\alpha^2(L) =  \langle  \hat{\mathbf{\Lambda}}^2 \rangle / L(L+1)$, with $\langle  \hat{\mathbf{\Lambda}}^2 \rangle$ given by Eq.~\eqref{ExpLambda}. Since in experiment the value of $B^\ast$ is usually determined from the splitting between the two lowest rotational states~\cite{ToenniesAngChem04}, we evaluate it from the first term of Eq.~\eqref{transH} for $L=1$:
\begin{equation}
\label{BstSC}
	\frac{B^\ast_\text{SC}}{B} = (1- \eta~\Delta)^2,
\end{equation}
where $\eta \equiv \alpha(1)/\Delta = \left(3 \sum_k f^2 (k)/\omega^2_{k} \right)^{1/2}$ is the only phenomenological parameter of the strong-coupling theory, which is obtained by fitting to the experimental data~\cite{sup}.

Figures~\ref{fig:BBstar}(a) and (b) show the comparison of Eq.~\eqref{BstSC} (red circles) with experiment (empty squares). We see that a good agreement with experiment is achieved for most molecules with $\gamma<1$:  SF$_6$, CS$_2$, HCCCN, OCS, N$_2$O, CO$_2$, CO, and NO. For I$_2$, the model overestimates the value of $B^\ast/B$ compared to the MC result~\cite{Shepperson16, ZillichI2} by about 20\%. It is worth noting that the calculations of impulsive alignment for I$_2$ in helium droplets performed with the MC value of $B^\ast/B=0.6$ predict the first revival later than observed in experiment~\cite{Shepperson16}, which signals that the experimental value of  $B^\ast/B$ might be larger than $0.6$.

For  C$_2$H$_2$, DCN, and HCN, see Fig.~\ref{fig:BBstar}(b), the disagreement with experiment is substantially larger. This fact might be due to the interplay between phonon and roton excitations in this region of rotational constants~\cite{ZillichHeC2H2, ZillichHeHCN, vonHaeftenHeCO}, which can potentially lead to a nontrivial dependence of the parameter $\eta$ on $B$ and $\Delta$. We note that for these three species classical hydrodynamics calculations lead to an underestimated $B^\ast/B$ ratio as well~\cite{CallegariPRL99, CallegariPRL99err}.  For LiH, Fig.~\ref{fig:BBstar}(b), strong renormalization of the rotational constant was previously predicted using path-integral MC simulations~\cite{ZillichHeLiH}. Here, due to a pronounced anisotropy of the He--LiH PES~\cite{TaylorHeLiH}, the strong-coupling angulon theory predicts $\langle  \hat{\mathbf{\Lambda}} \rangle > L$, which indicates the breakdown of the employed approximations. We attribute it to the fact that the PES features strong $U_1 (k)$ components which lead to processes involving simultaneous absorption or emission of  two phonons with $\lambda=1$, preserving the molecular angular momentum. The latter are not accounted for by the present theory. Nevertheless, Eq.~\eqref{BstSC} predicts a strong decrease of the LiH rotational constant in helium, which is in fair agreement with predictions of Ref.~\cite{ZillichHeLiH}. For light molecules of Fig.~\ref{fig:BBstar}(c), the results of the strong-coupling theory substantially deviate from experiment and are therefore not shown. 

Now let us consider the regime of weak coupling, $\gamma > 1$. There, one can obtain the energies of molecular states in helium using second-order perturbation theory on the Hamiltonian~\eqref{hamiltonian}:
\begin{equation}
\label{Epert}
	E_{L} = BL(L+1) - \sum_{k \lambda L' }  \frac{  U_\lambda(k)^2 \left[ C_{L 0, \lambda 0}^{L' 0} \right]^2}{B L'(L'+1) - BL(L+1) +\omega_k }
\end{equation}
where $C_{L 0, \lambda 0}^{L' 0}$ is the Clebsch-Gordan coefficient~\cite{Varshalovich}.  The second term of Eq.~\eqref{Epert} provides a shift which depends on the molecular rotational state, $L$, and therefore leads to renormalization of the rotational constant. It is interesting to note that the process described by Eq.~\eqref{Epert} -- differential shifts of the molecular rotational levels due to virtual phonon absorption -- represents an exact analogue of the  Lamb shift in quantum electrodynamics, which is induced by virtual photon excitations~\cite{ScullyZubairy, Rentrop16}.

Taking into account the dominant processes with $L=L'$ in Eq.~\eqref{Epert}, we obtain:
\begin{equation}
\label{BstPert}
	\frac{B^\ast_\text{WC}}{B} = 1- \xi \frac{\Delta^2 }{B},
\end{equation}
where $\xi= \sum_{k} f^2(k)/(5 \omega_k)$ is the only free parameter of the weak-coupling theory, obtained by fitting to the experimental data~\cite{sup}.

Figure~\ref{fig:BBstar}(c) compares Eq.~\eqref{BstPert} (blue triangles) with experimental data for light molecules. One can see that an agreement within 2\% is achieved for all the considered species: CH$_4$, CH$_3$, NH$_3$, HCl, H$_2$O, OH, and HF, which indicates the applicability of the weak-coupling angulon theory. We attribute a slightly larger disagreement for CH$_4$, NH$_3$, and H$_2$O to the approximation to the PES, Eq.~\eqref{U2}. We would like to point out that since the experiments on HCl~\cite{SkvortsovHeHCl} and OH~\cite{RastonHeOH} did not detect any significant  renormalization of the rotational constant, the corresponding experimental values of $B^\ast/B$ were set to $1$. While our theory indeed predicts $B^\ast/B \approx 1$ for the case of OH, we observe $B^\ast/B \approx 0.98$ for HCl, which is quite close to the corresponding value for HF. Given the similarities between the two species, we hope that this result will stimulate further measurements of $B^\ast/B$ for HCl.   For most heavy and  medium-mass species, Figs.~\ref{fig:BBstar}(a) and (b), the weak coupling theory fails to reproduce experimental data and is therefore not presented.

A peculiar situation occurs for three of the studied molecules, namely, C$_2$H$_2$, DCN, and HCN. There, the weak-coupling theory overestimates the $B^\ast/B$ ratio, while the strong-coupling approach underestimates it.  In principle, in order to obtain a quantitative agreement with experiment for these particular species, a different, intermediate-coupling angulon theory is required. However, as a rough approximation we can estimate the intermediate-coupling results by interpolating between the weak-coupling and strong coupling theories as $B^\ast_\text{IC} = (B^\ast_\text{SC}+B^\ast_\text{WC})/2$. The values of $B^\ast_\text{IC}/B$ are shown in Fig.~\ref{fig:BBstar}(b) by green crosses and are seen to provide a good agreement with experiment.

In addition to $B$-renormalization, previous experiments reported a significant increase in the centrifugal distortion constants, $D$, compared to the gas phase, obtaining e.g.\ $D=3.7 \cdot 10^{-5}$~cm$^{-1}$ for SF$_6$~\cite{HartmannHeSF6}, $D=1.6 \cdot 10^{-4}$~cm$^{-1}$ for HCCCN~\cite{CallegariHeHCCCN}, and $D=3.8 \cdot 10^{-4}$~cm$^{-1}$ for OCS~\cite{GrebenevOCS}. Such a distortion comes from the coupling between the angular momenta $\hat{\mathbf{L}}$ and $\hat{\mathbf{\Lambda}}$ in Eq.~\eqref{transH}~\cite{LehmannJCP01}. In order to obtain accurate values for $D$, a more involved, all-coupling angulon theory is required. Here we perform a rough estimate, based on second-order perturbation theory, which gives $D \sim \gamma^2/(5\xi)$~\cite{sup}. For the molecules mentioned above, we obtain $D \sim 4 \cdot 10^{-3},~1 \cdot 10^{-2}$, and $2 \cdot 10^{-2}$~cm$^{-1}$, respectively. Although this  estimate significantly exceeds the measured values, the drastic increase of $D$ in helium, as well as its qualitative change from molecule to molecule are in agreement with experiment.

In summary, we have demonstrated that the angulon theory is able to reproduce experimental data on the renormalization of rotational constants in superfluid $^4$He for 25 different molecules, based on only two phenomenological parameters. It has been shown that in the strong-coupling regime (mostly taking place for heavy and medium-mass molecules), the renormalization of molecular moments of inertia occurs due to a macroscopic deformation of the superfluid, which leads to redistribution of angular momentum between the molecule and excitations in helium. In the weak-coupling regime (applicable to  lighter species), the change in $B$ takes place due to a rotational Lamb shift induced by virtual single-phonon excitations.

These results provide strong evidence that molecules immersed  in superfluid $^4$He indeed form the angulon quasiparticles, and open the door for substantial simplifications of existing theories. As an example, the angulon theory is straightforward to apply to large polyatomic molecules and complexes studied in experiment~\cite{CallegariJCP01, ToenniesAngChem04, SzalewiczIRPC08} and can be extended to time-dependent problems of molecular dynamics in $^4$He~\cite{PentlehnerPRL13, PentlehnerPRA13, ChristiansenPRA15, Shepperson16}. Moreover,  the applicability of the angulon theory is not limited to bosonic quantum liquids. Therefore, it can potentially serve as a building block to understand molecular rotation in other types of solutions and solid-state environments.

We thank Gary Douberly, Bretislav Friedrich, and Igor Cherepanov for insightful discussions and Robert Zillich for sharing unpublished numerical results~\cite{ZillichI2}. This research was supported in part by the National Science Foundation under Grant No. NSF PHY11-25915.

\bibliography{droplets}

\newpage
\clearpage

\vspace{0cm}

\renewcommand{\thepage}{S\arabic{page}}  
\renewcommand{\thesection}{S\arabic{section}}   
\renewcommand{\thetable}{S\arabic{table}}   
\renewcommand{\thefigure}{S\arabic{figure}}
\renewcommand{\theequation}{S\arabic{equation}}

\setcounter{figure}{0} 
\setcounter{table}{0} 
\setcounter{equation}{0} 
\setcounter{page}{1} 

\onecolumngrid
\section{Supplemental Material}

\appendix

\section{The molecular anisotropy parameter $\Delta$}

We evaluate the anisotropy parameter as:

\begin{equation}
\label{DeltaPar}
	\Delta  = \frac{\vert V_\text{eff}^\parallel - V_\text{eff}^\perp\vert}{2} \sqrt{\frac{5}{4\pi}}
\end{equation}
 where $V_\text{eff}^\parallel $ and $V_\text{eff}^\perp$ give the average effective interactions for a helium atom in the parallel and perpendicular geometries, respectively. Table~\ref{delta} lists the values of $V_\text{eff}$, the corresponding molecule-helium geometries, as well as the resulting values of the anisotropy parameter $\Delta$.

\vspace{0.2cm}
\textit{For linear-rotor molecules,} $V_\text{eff}^\parallel$ is evaluated for the linear configuration of the molecule-helium complex, while $V_\text{eff}^\perp$ corresponds to a T--shaped configuration.  For the cases of a PES featuring potential minima (saddle points) both in the vicinity of the parallel and in the vicinity of the T--shaped  configurations, the values of $V_\text{eff}$ are set to the average depths of the corresponding minima (saddle points). If there is only one minimum (e.g. in the parallel configuration), the value of $V_\text{eff}^\perp$ is evaluated as the magnitude of the attractive potential at the same distance from the molecule as $V_\text{eff}^\parallel$. If, from one of the sides, the PES   is purely repulsive, the corresponding effective potential is set to zero, to reflect the fact that the helium density vanishes in this region. For open-shell molecules, such as OH$(X^2\Pi)$ and NO$(X^2\Pi)$ the symmetric (`sum') PES was used.

\vspace{0.3cm}
\textit{For symmetric and asymmetric tops,} $V_\text{eff}^\parallel$ is evaluated along the main symmetry axis of the molecule, while $V_\text{eff}^\perp$ is taken as an average potential in the direction perpendicular to the symmetry axis, laying within the plane of mirror symmetry. For example, for CH$_3$, $V_\text{eff}^\parallel$ is evaluated along the $C_3$ symmetry axis (where PES purely repulsive), while $V_\text{eff}^\perp$ is an average potential in the direction perpendicular to the $C_3$ axis, laying in the plane containing the C atom and one of the H atoms. For NH$_3$,  $V_\text{eff}^\parallel$ corresponds to the average minimum energy along the $C_3$ axis, while  $V_\text{eff}^\perp$ is an average potential perpendicular to the $C_3$ axis, laying in the plane containing the N atom and one of the H atoms. For H$_2$O, $V_\text{eff}^\parallel$ is an average potential along the $C_2$ symmetry axis,  while  $V_\text{eff}^\perp$ corresponds to the global minimum laying close to the direction perpendicular to the $C_2$ axis, within the mirror symmetry plane (the one containing the O atom and bisecting the H--H bond).

 \vspace{0.3cm}
\textit{For  spherical tops,} the anisotropy parameter $\Delta$ was evaluated as an average anisotropy between the potential minima along the main symmetry axes of the molecule-helium system. That is,   $\vert V_\text{eff}^\parallel - V_\text{eff}^\perp\vert$ in Eq.~\eqref{DeltaPar} was replaced by $\sum_{i\neq j} \vert V_\text{eff}^{(i)} - V_\text{eff}^{(j)}\vert$, where $i,j$ label all non-equivalent minima of the PES. In the case of SF$_6$, for example, $V_\text{eff}^{(i,j)}$   represent the  minima along the $C_4$, $C_3$, and $C_2$ axes (so-called `vertex,' `face,' and `edge' geometries, respectively), while for CH$_4$ we included two minima along the $C_3$-axis (`vertex' and `face') and one minimum along the $C_2$-axis (`edge').

\begin{table*}[h]
{\footnotesize
\caption{Effective molecule-helium interaction in parallel and perpendicular geometries, $V_\text{eff}^\parallel$ and $V_\text{eff}^\perp$ (`av.' labels $V_\text{eff}$ obtained through averaging), the resulting anisotropy parameter, $\Delta$, as well as the dimensionless coupling constant $\gamma$. For spherical tops, $\Delta$ was calculated as an average anisotropy of the PES (see text).  The experimental values for renormalization of the rotational constant, $B^\ast_\text{exp}/B$, are listed as well, where $^\ast$ marks the results of Monte-Carlo simulations.  \label{delta} \vspace{0.5cm}}
\begin{tabular}{| c | c | c | c | c | c | c |}
         \hline
   \multicolumn{7}{| c|}{\bf Linear rotors} \\
     \hline
 molecule & $V_\text{eff}^\parallel [\text{cm}^{-1}]~(r \text{[\AA]}, \theta [^\circ])$ & $V_\text{eff}^\perp [\text{cm}^{-1}]~(r \text{[\AA]}, \theta [^\circ])$ & $\Delta  [\text{cm}^{-1}]$  & $\gamma $ & $B^\ast_\text{exp}/B$ & Ref.   \\[3pt]
  \hline
         CS$_2$ & $-30.8~(5.0, 0)$  & $-53.0~(3.4, 90)$ &  6.97  & 0.0156  & 0.3$^\ast$ & \cite{ZangJCP14, ZillichI2}  \\
             I$_2$ & $-45.2~(4.9, 0)$  &  $-37.7~(3.8, 90)$ &  2.35 & 0.0159 &  0.6$^\ast$ & \cite{KalemosJCP12, GutierrezHeI2, Shepperson16}  \\
   	HCCCN &  $-26.6~[\text{av.}~(5.6, 0), (5.3, 180)]$ &  $-48.7~(3.3, 94)$  &  6.97 & 0.0217 & 0.35 & \cite{CallegariHeHCCCN, AkinOjoJCP03}  \\
    	    OC$^{34}$S &   $-29.5~[\text{av.}~(4.7, 0), (4.4, 180)]$  &  $-50.2~(3.3, 70)$ & 6.53   & 0.0303  & 0.36 & \cite{HowsonHeOCS, GrebenevOCS}  \\
		    OC$^{32}$S &  --  &  --  & --  & 0.0311 &0.36 & --  \\
                $^{14}$N$^{15}$NO  &  $-30.6~[\text{av.}~(4.1, 0), (4.5, 180)]$  & $-65.7~(3.0, 90)$ & 11.1 & 0.0366  & 0.17 & \cite{ChangN2O, NakautaCO2N2O}\\
                         $^{14}$N$^{14}$NO  &  --  & --  & --  & 0.0378  & 0.17 & -- \\
  	CO$_2$  &  $-26.7~(4.3, 0)$&   $-49.4~(3.1, 90)$& 7.16 & 0.0545 & 0.39 & \cite{RanHeCO2, NakautaCO2N2O}\\
	      LiH & $-90~[\text{av.}~(2.3, 0), (4.9, 180)]$  &  $-15.0~(4.2, 90)$ & 23.6 & 0.317 & 0.06$^\ast$ &  \cite{TaylorHeLiH, ZillichHeLiH}\\ 
	   DCN & $-25.8~[\text{av.}~(4.2, 0), (4.2, 180)]$  &  $-14.0~(4.2, 90)$ &  3.70  &  0.326   & 0.83 & \cite{HaradaHeHCN, ConjusteauHeHCN, ZillichHeHCN} \\
                                HCN & --   &  -- &  -- &  0.399   & 0.81 & -- \\
    $^{13}$C$^{18}$O & $-10.0~[\text{av.}~(3.4, 0), (3.4, 180)]$ &  $-22.3~(3.4, 71)$&  3.89 & 0.447& 0.63 &  \cite{PetersonJCP05, HeijmenHeCO, vonHaeftenHeCO}   \\ 
        $^{12}$C$^{18}$O & --  &  -- &  --  & 0.468 & 0.63 & --   \\ 
               $^{13}$C$^{16}$O & --  &  -- &  --  & 0.470 & 0.63 & --   \\ 
    $^{12}$C$^{16}$O & --  &  -- &  --  & 0.491 & 0.63 &  --   \\     
       C$_2$H$_2$ & $-24.2~(4.3, 0)$  &  $-16.7~(3.8, 90)$ & 2.37 & 0.494 & 0.88 & \cite{FernandezHeC2H2, NautaHeC2H2, ZillichHeC2H2}\\
              NO & $-19.0~[\text{av.}~(3.9, 0), (3.9, 180)]$  &  $-25.0~(3.3, 90)$ &  1.89  & 0.874 & 0.76 & \cite{KlosHeNO, vonHaeftenHeNO}  \\    
     HCl & $-31.5~[\text{av.}~(3.9, 0), (3.4, 180)]$  &  $-17.1~(3.8, 90)$  &  4.54 & 2.23 & $\sim 1$ &  \cite{MurdachaewHeHCL, SkvortsovHeHCl}\\
      HF & $-36.8~[\text{av.}~(3.2, 0), (2.9, 180)]$  &  $-17.5~(3.2, 90)$ &  6.07 & 3.26 & 0.98  & \cite{MoszynskiHeHF, NautaHeHF}\\   
     OH & $-20.0~[\text{av.}~(3.4, 0), (3.4, 180)]$  &  $-16.0~(3.4, 90)$ &  1.26 & 14.7 & $\sim 1$ & \cite{PatersonHeOH, RastonHeOH}\\ 
     \hline
     \multicolumn{7}{c}{ }\\[5pt]
         \hline
   \multicolumn{7}{| c|}{\bf Symmetric and asymmetric tops} \\
        \hline
 & $V_\text{eff}^\parallel [\text{cm}^{-1}]~(r \text{[\AA]}, \theta [^\circ], \phi [^\circ])$ & $V_\text{eff}^\perp [\text{cm}^{-1}]~(r \text{[\AA]}, \theta [^\circ], \phi [^\circ])$ &   \multicolumn{3}{c}{}   &  \\[3pt]
  \hline
     \hline
          NH$_3$ & $-9.68~[\text{av.}~(3.9, 0, 0), (4.1, 180, 0)]$  &  $-28.0~[\text{av.}~(3.2, 90, 60), (3.6, 85, 0)]$  &  5.78  & 1.71 & 0.95  &\cite{HodgesHeNH3, SlipchenkoHeNH3}\\
       CH$_3$ & $0~(3.4, 0, 0)$ &  $-13.5~[\text{av.}~(3.4, 0, 0), (3.4, 90, 60)]$  & 4.26 & 2.22 & 0.98  & \cite{DagdigianHeCH3, MorrisonJPCA13}\\ 
   H$_2$O    &  $-21.6~[\text{av.}~(3.3, 0, 0), (3.5, 180, 0)]$ & $-34.9~(3.1, 75, 0)$ &  4.22 & 2.82  & 0.97 &  \cite{PatkowskiHeH2O, KuyanovHeH2O}\\    
         \hline
        \multicolumn{7}{c}{ }\\[5pt]
         \hline
   \multicolumn{7}{| c|}{\bf Spherical tops} \\
           \hline
  & \multicolumn{2}{|c|}{$V_\text{eff} [\text{cm}^{-1}]~(r \text{[\AA]}, \theta [^\circ], \phi [^\circ])$}   &      \multicolumn{3}{c}{}     &   \\[3pt]
  \hline
     \hline
         SF$_6$ &  \multicolumn{2}{|c|}{ $-57.7~(3.9, 54.7, 45)$; $-43.1~(4.2, 90, 45)$; $-27.8~(4.5, 0, 0)$ }  &  6.28 & 0.0145 & 0.37 & \cite{BarnettHeSF6, PackHeSF6, HartmannHeSF6, LeeHeSF6}  \\
            CH$_4$  & \multicolumn{2}{|c|}{ $-36.0~(3.3, 54.7, 90)$; $-25.3~(3.6, 0, 0)$; $-17.8~(4.0, 54.7, 0)$ }    &  3.83 &    1.37  & 0.96   &  \cite{CalderoniHeCH4, NautaHeCH4}\\  
\hline
\end{tabular}
}
\end{table*}

\clearpage

\section{Fitting procedure to obtain $\eta$ and $\xi$}

\begin{figure}[h]
\centering
\includegraphics[width=0.4\textwidth]{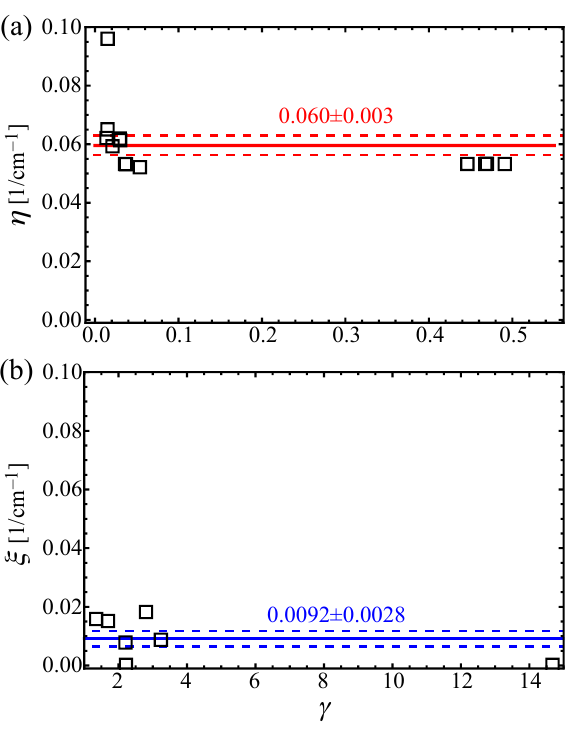}
\caption{(a) The values of $\eta$ extracted from the experimental data for molecules interacting with helium strongly ($\gamma<1$): CS$_2$, I$_2$, SF$_6$, HCCCN, OCS, N$_2$O, CO$_2$, and CO (empty squares), along with the results of the the linear fit (red lines). (b)~The values of $\xi$ extracted from the experimental data for molecules interacting with helium weakly ($\gamma>1$):  HF, HCl, NH$_3$, CH$_3$, CH$_4$, H$_2$O, and OH (empty squares), along with the results of the the linear fit (blue lines).}
\label{fig:fitting}
\end{figure}

We obtain the strong-coupling and weak-coupling constants of the model, $\eta$ and $\xi$, by linear fitting to the experimental data  from Refs.~\cite{ZillichI2, Shepperson16, AkinOjoJCP03, GrebenevOCS, NakautaCO2N2O,  ConjusteauHeHCN, ZillichHeHCN,  ZillichHeLiH, vonHaeftenHeCO, NautaHeC2H2, ZillichHeC2H2, vonHaeftenHeNO, SkvortsovHeHCl, NautaHeHF, RastonHeOH,  SlipchenkoHeNH3, MorrisonJPCA13, KuyanovHeH2O, HartmannHeSF6, LeeHeSF6, NautaHeCH4}. Fig.~\ref{fig:fitting} shows the values of $\eta$ and $\xi$ extracted from the experimental data using  Eqs.~(10) and~(12) of the manuscript, respectively, and the values of $\Delta$ from Table~\ref{delta} (empty squares). The horizontal lines indicate the best fit with the corresponding error bar.

\clearpage
\section{The centrifugal distortion constant $D$}

The strong-coupling states of Eq.~(8) represent exact eigenstates of Eq.~(7) only in the limit of a slowly-rotating molecule, $B \to 0$. For $B \neq 0$, the coupling between the angular momentum vectors $\hat{\mathbf{L}}$ and $\hat{\mathbf{\Lambda}}$ will change the form of the many-particle state. In turn, it will lead to additional terms in the angulon rotational energy, $\Delta E_{L} = - D [L(L+1)]^2 + \dots$, where $D$ is the so-called `centrifugal distortion constant.'

In principle, in order to obtain a numerically accurate value of $D$, one would need to develop an all-coupling theory based on the Hamiltonian~(7). We can, however, estimate the next-order correction to the angulon energy applying second-order perturbation theory to the strong-coupling states~(8). Let us calculate the effect arising from the coupling of the states~(8) to all possible non-interacting molecular states, $\ket{L'M'} \ket{0}$, due to the first term of the Hamiltonian~(7):
\begin{equation}
\label{DeltaE}
	\Delta E_{LM} \sim - \sum_{L'M'} \frac{\vert \bra{0} \bra{L' M'}  B (\hat{\mathbf{L}} - \hat{\mathbf{\Lambda}})^2  \ket{\psi_{LM}}  \vert^2 }{E_{LM}},
\end{equation}
where $E_{LM} = - \sum_{k \lambda}  U^2_\lambda (k)/\omega_{k}$ are the energies of the dressed states $\ket{\psi_{LM}}$. In terms of our model parameters, the energy correction \eqref{DeltaE} assumes the following form:
\begin{equation}
\label{DeltaE2}
	\Delta E_{L} \sim - \frac{B^2}{5 \xi \Delta^2} e^{-\eta^2  \Delta^2/3 } \left[ L(L+1) \right]^2
\end{equation}
For the molecules listed in Table~\ref{delta}, the exponential factor provides a negligible contribution. Therefore, the centrifugal distortion constant can be approximated as:
\begin{equation}
\label{Dconst}
	D \sim \frac{\gamma^2}{5 \xi} 
\end{equation}

\end{document}